\documentclass{JHEP3}
\usepackage{epsfig}
\usepackage{graphicx}
\usepackage{enumerate}
\usepackage{amsmath}

\def\ben{\begin{equation}}
\def\een{\end{equation}}
\def\bena{\begin{eqnarray}}
\def\eena{\end{eqnarray}}

\title{Asymptotic Flatness, Little String Theory, and Holography}

\author{Donald Marolf  \\
Physics Department, UCSB, Santa Barbara, CA 93106
\texttt{marolf@physics.ucsb.edu}}

\abstract{We argue that any non-gravitational holographic dual to
asymptotically flat string theory in $d$-dimensions naturally
resides at spacelike infinity.  Since spacelike infinity can be
resovled as a $(d-1)$-dimensional timelike hyperboloid (i.e., as a
copy of de Sitter space in $(d-1)$ dimensions), the dual theory is
defined on a Lorentz signature spacetime.  Conceptual issues
regarding such a duality are clarified  by comparison with linear
dilaton boundary conditions, such as those dual to little string
theory.  We compute both time-ordered and Wightman boundary 2-point
functions of operators dual to massive scalar fields in the
asymptotically flat bulk.}

\date{November, 2006}


\preprint{}

\begin{document}


\section{Introduction}
\label{intro}

The discovery of gauge/gravity dualities \cite{Juan,BFSS} has
profoundly influenced string theoretic investigations of quantum
gravity.  In contexts where they are known, these dualities appear
to provide a complete non-perturbative formulation of the
theory and allow one to study the emergence of (bulk) spacetime as
an effective description when curvatures are small; see e.g.
\cite{DB}.    If this program is fully successful, one will be able
to address deep questions concerning black holes, singularities, and
the like by performing definite calculations in the gauge theory
dual.

However, such dualities are known only when certain boundary
conditions are imposed on the bulk string theories.  The best
studied case is that of asymptotically anti-de Sitter (AdS) boundary
conditions (crossed with some compact manifold), and other
well-studied examples \cite{IMSY,JP,ABKS}  have qualitatively similar
boundary conditions. It is clearly of interest to understand if
dualities exist in more general settings.  The case of
asymptotically de Sitter spacetimes has received much attention (see
e.g. \cite{dSCFT1,dSCFT2}) and some simple cosmologies have been
studied, but asymptotically flat settings are relatively unexplored.

We will pursue the asymptotically flat context here.  At first
glance, it might appear that an asymptotically flat holographic
duality must differ radically from AdS/CFT.  Theories dual to AdS
are associated with the $S^{n-1} \times {\mathbb R}$ which forms the
conformal boundary of AdS, and which is a Lorentz-signature
spacetime in its own right. In contrast, the smooth part (${\cal
I}$) of the asymptotically-flat conformal boundary is well-known to
be null. This makes it more difficult to imagine ${\cal I}$ as a
home for a dual theory, though several interesting attempts have
been made \cite{dBS,ADAF,ACH}.
\FIGURE{\includegraphics[width=4in]{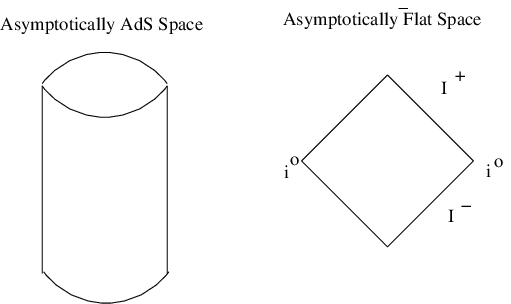}\caption{Conformal
diagrams of asymptotically AdS space (on the left) and
asymptotically flat space (on the right).  The two corners marked
$i^0$ on the asymptotically flat diagram are to be identified.}
\label{AdSAF}}

However, as emphasized in \cite{Witten}, the notion of a conformal boundary  is itself not fundamental to AdS/CFT. Rather, this role derives from the
convenient way in which the AdS conformal boundary parametrizes the
space of possible boundary conditions on propagating fields. Each
possible boundary condition defines a bulk theory, which is then dual to
a particular non-gravitating field theory (with a particular
Lagrangian) on $S^{n-1} \times {\mathbb R}$.  In contrast, in the
asymptotically flat setting, data on the conformal boundary (${\cal
I}$) naturally encodes (part of) the information about the {\it
state};  changing a solution on ${\cal I}^-$ (${\cal
I}^+$) alters the initial (final) data but does not change the
dynamics. Instead, boundary conditions are naturally imposed at {\it
spacelike} infinity, $i^0$.  This fact will be reviewed in detail in
section \ref{bcs} below.

Although $i^0$ is represented by a single point in the conformal
compactification, it may be better thought of as a timelike
hyperboloid.   A particularly nice construction of this `boundary'
was given in \cite{AR}.  The essential point is that physical fields
do not admit smooth limits at the point $i^0$ of the conformal
diagram. Instead, they admit limits which depend (smoothly) on the
spacelike direction along which one approaches $i^0$ (see e.g.
 \cite{AshtekarHansen, AshtekarInf}).  Thus, for a $d$-dimensional asymptotically flat
spacetime, the asymptotics (and, as we will see, the boundary
conditions) are associated with functions on the $(d-1)$-dimensional
hyperboloid ${\cal H}$ of spacelike directions.  Note that ${\cal
H}$ is naturally regarded as a signature $(d-2)+1$ manifold, and
that it is isometric to the unit $(d-1)$-dimensional de Sitter
space.  One imagines that ${\cal H}$ may provide a more hospitable
home for a dual theory than ${\cal I}$.

The above reasoning is strengthened by a comparison with linear
dilaton backgrounds.  Despite certain complicating features,  string
theory in appropriate linear dilaton backgrounds is known to be dual
to a (in this case, non-local) non-gravitating theory \cite{ABKS}.
The prime example is the case of little string theory \cite{NSNS},
where string theory with asymptotics given by the near-horizon limit
of $N$ NS5-branes is dual to the low-energy limit of the open string
theory on the branes. Here the ground state of the bulk theory is
described by the string-frame metric
\begin{equation}
ds^2_{string} = dx_6^2 + dz^2 + d \Omega_3^2,
\end{equation}
where $dx^2_6$ is the 5+1 Minkowski metric, and $d \Omega_3^2$ is
the metric on the unit-sphere.  Due to the $dx_6^2 + dz^2$ factor,
it is clear that the smoothest part of the conformal boundary is
just ${\cal I}^6$, the null boundary of 6+1 Minkowski
space\footnote{Even this boundary is not strictly smooth due to the
fact that conformal compactification shrinks the 3-sphere to zero
size on the boundary.}. However,  the non-gravitating dual lives on
the 5+1 Minkowski space ${\mathbb{R}}^{5,1}$ associated with the branes, and not on
${\cal I}^6$. The ${\mathbb{R}}^{5,1}$ is a Lorentz signature spacetime which, as
described in \cite{ABKS}, can be associated with the large $z$
asymptotics of the linear dilaton spacetime.  Roughly speaking, the
dual little string theory ``lives'' at spacelike infinity.  This is
exactly what we propose in the asymptotically flat context.

Some of the above reasoning was used in \cite{MM,MMV} to motivate
the introduction of a (classical) boundary stress tensor on ${\cal
H}$.  The stress tensor is a one-point function, and our interest
here will be in generalizing the discussion to both the quantum case
and to higher ($n$-point) boundary correlators.  We begin by
reviewing the structure of fields near spatial infinity and
discussing possible (infinitesimal) deformations of the usual
boundary conditions in section \ref{bcs}. Section \ref{formal} then
considers variations of the path integral with respect to these
boundary conditions.  First variations lead to a natural definition
of boundary operators. However, subtleties arise for higher
correlators.

Section \ref{concrete} computes boundary two-point functions for
operators dual to both massive and massless bulk fields.  Two
different computations are considered.  The first uses the on-shell
action and attempts to apply the method used by Gubser, Klenanov,
and Polyakov  in the AdS context \cite{GKP}. This method, however,
does not appear to give a useful answer to our problem. Instead,
non-local analogues of `contact terms' make the result ambiguous.
This same feature previdents a straightforward application of the
method of \cite{Witten}. However, our second computation is more
successful. Here we first calculate the boundary Wightman function,
which is free of contact divergences and well-defined. The result
then determines the time-ordered two-point function and guarrantees
that it has the expected analytic structure. For comparison, we show
that similar results hold for linear dilaton backgrounds dual to
little string theory.   We close with some further discussion in
section \ref{disc}.

\section{Fields near spatial infinity}
\label{bcs}

The complete definition of a field theory generically requires a
choice of boundary conditions. In finite volume or in asymptotically
anti-de Sitter spacetimes, the fact that signals can propagate from
the boundary to the bulk makes the need for boundary conditions
especially clear.    Boundary conditions are required to fully
specify the evolution, as well as to conserve symplectic flux (and
thus to make any covariant phase space well-defined).

Boundary conditions in asymptotically flat space are more subtle,
but their necessity can be clearly seen in the context of, e.g.,
non-gravitating scalar field theories on Minkowski space.  The scalar
wave equation admits many solutions which diverge at spatial
infinity, and which cannot be allowed as propagating degrees of
freedom if either the energy or the symplectic structure is to be
finite.  In order to construct a well-defined phase space, one must
{\it fix} the part of the field associated with such
(non-normalizable) modes, allowing only the normalizable part to be
dynamical.

The fixed non-normalizable modes are a background structure which
play the same roles as do boundary conditions in finite volume.   In
general, it is necessary only to fix the asymptotic behavior of the
field.  As a result, we may think of their specification as
corresponding to a choice of ``boundary conditions at spatial
infinity,'' and we will use this terminology below.  Much the same association between boundary conditions and
non-normalizable modes is familiar in the anti-de Sitter context
(see e.g. \cite{Witten,GKP,BLK,BLKT}).

Many readers may think of spatial infinity as a single point ($i^0$)
in the Penrose compactification (see e.g. \cite{HE}) of Minkowski
space. However, for the reasons stated in section \ref{intro}, it is better to consider spacelike infinity to be the
hyperboloid ${\cal H}$ (see \cite{AR,AshtekarHansen, AshtekarInf})
of spacelike directions.  To understand this description, recall that the line element of
$d$-dimensional Minkowski space may be written in hyperbolic
coordinates as
 \begin{equation}
 \label{hyp}
 ds^2 = d \rho^2 + \rho^2 \omega_{ij} d\eta^i d\eta^j,
 \end{equation}
where $\rho^2 = x_ax^a$, $\omega_{ij}$ is the metric on the unit
$(d-1)$-dimensional Lorentz-signature hyperboloid ${\cal H}$, and
$\eta^i$ are coordinates on ${\cal H}$.  Spacelike infinity is
essentially the large $\rho$ limit of the constant $\rho$
hyperboloids.

\subsection{Massive Free Scalars}
\label{massive}

Let us recall the structure of solutions to the (massive) free
scalar wave equation on Minkowski space in the hyperbolic
coordinates (\ref{hyp}).  The wave equation is
\begin{equation}
\label{boxphi}
 0 = (\Box - m^2) \phi = \left( \frac{1}{\rho^{d-1}}
\partial_\rho \rho^{d-1} \partial_\rho + \frac{1}{\rho^2}
\nabla^2_{\cal H} - m^2\right) \phi,
\end{equation}
where $\nabla^2_{\cal H}$ is the scalar D'Alembertian on ${\cal
H}$ and $m^2 > 0$.  Any solution to (\ref{boxphi}) is a linear combination of
the modes
\begin{eqnarray}
\label{mmodes}
 \Phi_{q, \vec j} & =&   \rho^{\frac{2-d}{2}} \hat I_\nu(m \rho)    Y_{q,\vec j}, \ \ \ {\rm and} \cr
\tilde \phi_{q, \vec j} &=& \rho^{\frac{2-d}{2}}  K_\nu(m \rho)    Y_{q,\vec j},
\end{eqnarray}
where  $ I_\nu(m \rho),  K_\nu(m \rho) $ are the usual modified
Bessel functions with $\nu = \sqrt{-q^2 +
\left(\frac{d-2}{2}\right)^2}$ and $\hat I_\nu$ denotes the real
part\footnote{Recall that $K_\nu$ is real for all real $\nu^2$, but
that $I_\nu$ is real only $\nu^2 > 0$.  It is useful to choose our
mode functions to be real for $\nu^2 <0$ as well.} of $I_\nu$. Thus,
$\hat I_{\nu} = \hat I_{-\nu}$ for imaginary $\nu$.    The $
Y_{q,\vec j}$ are harmonics on ${\cal H}$ satisfying
\begin{equation}
\nabla_{\cal H}^2 Y_{q,\vec j} = q^2 Y_{q,\vec j}.
\end{equation}
At least for normalizable modes, we choose $Y_{q, \vec j}$ so that
$K_\nu(m \rho) Y_{q, \vec j}$ is purely positive (negative)
frequency in the usual sense on Minkowski space, according to the
positive (negative) sign of $q$.  Such a choice must be possible
since any normalizeable mode has a unique decomposition into
positive and negative frequency parts, and the decomposition
respects Lorentz invariance.  Thus, the `projection' onto the
positive frequency subspace commutes with each element of the
Lorentz group and is proportional to the identity in any irreducible
representation.

The index $\vec j$ is an additional label to account for all further
degeneracies.  For later use, we note that $\vec j$ specifes a
harmonic $Y_{\vec j}$ on $S^{d-2}$.  Thus, $\vec j$ specifes an
integer spin representation of $SO(d-1)$, as well as a state within
this representation; e.g, $\vec j = (j,m)$ for $SO(3)$. We denote
the usual quadratic Casimir of $SO(d-1)$ by $|\vec j|^2 = j(j + d
-3)$ for $j \in \mathbb{Z}$.

Let us ask which solutions above represent propagating degrees of
freedom.  First, propagating modes should be (Klein-Gordon) normalizable at large
$\rho$, restricting them to linear combinations of the $\tilde
\phi_{q, \vec j}$.  Second, they should be
normalizable at small $\rho$.  However, for real $\nu$ the Bessel
function $K_\nu$ grows like $\rho^{-\nu}$ as $\rho \rightarrow 0$.  Thus,
normalizeable modes satisfy
 \begin{equation}
 \label{qbound}
  q^2 \ge   \left( \frac{d-2}{2} \right)^2.
  \end{equation}
As noted in, e.g. \cite{tagirov1,tagirov2}, such $Y_{q, \vec j}$ lie
in the princpal series of $SO(d-1,1)$ representations, except for
the marginal case $q = \pm \frac{(d-2)}{2}$ (which is  a member of
the complimentary series, see e.g. \cite{TT2}). As a result, they
are delta-function normalizable in $L^2({\cal H})$. We take them to
satisfy
 \begin{equation}
 \label{Ynorm}
 \int_{\cal H} \sqrt{-\omega} \ Y^*_{q, \vec j} Y_{q', \vec j'} =
 \delta(q-q') \delta_{\vec j, \vec j'}.
 \end{equation}

The $L^2({\cal H})$ normalizability of $Y_{q, \vec j}$ corresponds
to the expected behavior of propagating fields in the distant future
and past along the hyperboloid ${\cal H}$ as follows:  In the
distant past and future, each constant $\rho$ hyperboloid ${\cal
H}_\rho$ approaches the null cone through the origin. Now, solutions
to (\ref{boxphi}) decay along this null cone in the same manner as
the massive Green's function (as $r^{-(d-1)/2}$).  However, the
volume of spherical slices of ${\cal H}$ grows only as $r^{d-2}$.
Since $r$ increases exponentially with proper time along ${\cal H}$,
we see that smooth solutons should lie in each $L^2({\cal H}_\rho)$.
 In general, normalizable solutions can be obtained through
(continuous) superpositions of modes normalized as in (\ref{Ynorm}).

Thus, the modes $\tilde \phi_{q, \vec j}$ (satisfying
(\ref{qbound})) form a basis for the propagating solutions.  Other
modes which have divergent Klein-Gordon norm at large $\rho$ are
non-dynamical and must describe a fixed background.  Such modes are
specified as part of the  ``boundary condition'' which defines the
system.  We see that any $\Phi_{q, \vec j}$ provides a well-defined
boundary condition of this sort.  However, such
boundary conditions will be of less interest if they are orthogonal to
all normalizable modes with respect to the inner product on
$L^2({\cal H})$.  Since the $Y_{q, \vec j}$ are eigenstates in
$L^2({\cal H})$ of the self-adjoint operator $\nabla^2_{\cal H}$
with eigenvalue $q^2$, and since eigenstates with different
eigenvalues are orthogonal, the most interesting boundary conditions will
also satisfy (\ref{qbound}). As a result, we will often restrict
consideration below to those $\Phi_{q, \vec j}$ which satisfy
(\ref{qbound}).

Let us write a general solution as $\phi = \Phi + \tilde \phi$,
where $\Phi$ is a superposition only of the $\Phi_{q, \vec j}$ modes
and $\tilde \phi$ is a superposition only of the modes $\tilde
\phi_{q, \vec j}$, both satisfying (\ref{qbound}).  Asymptotically
we have
\begin{eqnarray}
\label{IKasympt}
I_\nu(m \rho) \sim   \frac{e^{m \rho}}{\sqrt{2 \pi m \rho}}, \cr
K_\nu(m \rho) \sim  \frac{e^{-m \rho}}{\sqrt{2  m \rho/ \pi}},
\end{eqnarray}
so that $\Phi,\tilde \phi$ can be characterized by arbitrary
functions $\alpha, \beta$ on ${\cal H}$ defined  by
\begin{eqnarray}
\label{ab} \alpha (\eta) &:=& 2m \lim_{\rho \rightarrow \infty}
\rho^{\frac{d}{2}} K_\nu(m \rho) \Phi(x) = \lim_{\rho \rightarrow
\infty} (\rho^{\frac{d}{2}} K_\nu(m \rho)) \overleftrightarrow
\partial_\rho \Phi , \cr
\beta (\eta) &:=& - 2m \lim_{\rho \rightarrow \infty}
\rho^{\frac{d}{2}} I_\nu(m \rho) \tilde \phi(x) = \lim_{\rho
\rightarrow \infty} (\rho^{\frac{d}{2}} I_\nu(m \rho))
\overleftrightarrow
\partial_\rho \tilde \phi .
\end{eqnarray}
From (\ref{IKasympt}) we see that (\ref{ab}) is independent of the
choice of $\nu$.  Boundary conditions which require the full
solution $\phi  = \Phi + \tilde \phi$ to be normalizable will be
called ``fast fall-off'' boundary conditions; these clearly impose
$\alpha =0$.

Equation (\ref{ab}) describes a natural pairing between boundary
conditions and propagating solutions.  It is useful to write this pairing in
terms of the ``boundary product'' ($\circ$) of two solutions
$\phi_1= \Phi_1 + \tilde \phi_1, \phi_2= \Phi_2 + \tilde \phi_2$:
\begin{equation}
\label{pair} \phi_1 \circ \phi_{2} = \int_{\partial {\cal M}}
\sqrt{-h} \ \phi_1 n^a \overleftrightarrow \partial_a \phi_{2} =
\int_{\cal H} \sqrt{-\omega}  (\alpha_1 \beta_2 - \alpha_2 \beta_1),
\end{equation}
where $\alpha_i, \beta_i$ are defined as in (\ref{ab}) using
$\Phi_i, \tilde \phi_i$ respectively.  The symbol $\int_{\partial
{\cal M}}$ denotes the $\rho \rightarrow \infty$ limit of a family
of integrals, each performed over a hyperboloid ${\cal H}_\rho$ at
fixed $\rho$ having unit outward-pointing normal $n^a$ and induced
metric $h_{ij} = \rho^2 \omega_{ij}$.  As argued above, one expects
propagating modes to have $\beta \in L^2({\cal H})$. Clearly, it is
also natural to take $\alpha \in L^2({\cal H})$, in which case
(\ref{pair}) is finite.

\subsection{Massless Free Scalars}
\label{massless}

Let us now consider the special case $m=0$.  In this limit the mode
functions are no longer exponential at infinity, and their
asymptotic behavior now depends on the harmonic on ${\cal H}$.  Any
solution to the massless Klein-Gordon equation is a linear
combination of the modes $ \rho^{\lambda_\pm} Y_{q,\vec j}$, where
\begin{equation}
\lambda_{\pm} = - \frac{d-2}{2} \pm \sqrt{  \left( \frac{d-2}{2}
\right)^2 - q^2}.
\end{equation}
As in the massive case, normalizability considerations require
propagating modes to be oscillatory near $\rho=0$, and so again
impose (\ref{qbound}).

Now, when (\ref{qbound}) holds, the massless modes are also
oscillatory at large $\rho$.  Thus, there is some freedom with
regard to which modes are considered to be dynamical and which modes
are taken to define boundary conditions. One may check, however,
that for the system to have a well-defined phase space (and, in
particular, for the symplectic flux through ${\cal H}$ to vanish), that
one may allow only a single propagating mode for each $(q, \vec j)$
satisfying (\ref{qbound}).  The situation appears to be analogous to
that of scalars in AdS with masses close to the
Breitenlohner-Freedman bound \cite{BF}, in which there is a large
freedom to choose boundary conditions. We shall assume that some
particular choice has been made and denote the corresponding
propagating modes by $\tilde \phi_{q, \vec j}$.  We denote another
linearly independent set of modes by $\Phi_{q, \vec j}$, which we
think of as describing particularly simple boundary conditions;
namely, those boundary conditions for which the space of propagating
modes remains unchanged from the choice made above.  We require only
that $\Phi_{q, \vec j}, \tilde \phi_{q, \vec j}$ are each of the
form $f(\rho) Y_{q, \vec j}$ and that our modes satisfy the
normalization conditions:

\begin{eqnarray}
\label{mlessnorms}
 \Phi^*_{q,' \vec j'} \circ \tilde \phi_{q, \vec j} &=& -\delta(q-q') \delta_{\vec j, \vec j'}, \cr
 \langle \tilde \phi^*_{q, \vec j},\tilde \phi_{q', \vec j'} \rangle_{KG}
 &:=& i \int_\Sigma \sqrt{g_\Sigma} \ \tilde \phi^*_{q, \vec j} n^a \overleftrightarrow \partial_a \tilde \phi_{q', \vec j'}
 = {\rm sign} (q) \frac{\mu }{2} |\Gamma(i\mu)|^2 \delta(q - q') \delta_{\vec j, \vec j'},
\end{eqnarray}
where $\mu = \sqrt{q^2 - \left(\frac{d-2}{2}\right)^2}$ and $\Sigma$
is a Cauchy surface with induced volume element  $\sqrt{g_\Sigma}$
and unit future-pointing normal $n^a$.   We have chosen the normalization factor on the
right-hand side of (\ref{mlessnorms}) in order to mirror the
normalization of the massive modes (which is computed in the
appendix, see (\ref{mnorms})).

For $m^2 > 0$, we parametrized the linear solutions in terms of two
functions $\alpha, \beta$ on ${\cal H}$. This is again possible
here; for example, one may take
\begin{eqnarray}
\label{mlessab}
 \alpha &=& \sum_{\vec j} \int_{q^2 \ge \left((d-2)/2\right)^2 } dq \   Y_{q, \vec j} \ \tilde \phi^*_{q,
\vec j} \circ \Phi \ \ \ {\rm and} \cr
 \beta &=&   \sum_{\vec j}
\int_{q^2 \ge \left((d-2)/2\right)^2 } dq \ Y_{q, \vec j} \
\Phi^*_{q, \vec j} \circ \tilde \phi,
\end{eqnarray}
where $\circ$ again denotes the boundary product (\ref{pair}). As before, $\tilde \phi$ denotes a general linear combination of the $\tilde \phi_{q,
\vec j}$, and $\Phi$ denotes a general linear combination of the
$\Phi_{q, \vec j}$.  Equations (\ref{mlessab}) are the natural extension to $m^2=0$ of
$\alpha, \beta$ defined in (\ref{ab}) for $m^2 > 0$.  However, for massless fields it is less clear that (\ref{mlessab}) gives a natural notion of locality on ${\cal H}$.  We see that the extraction of $\alpha, \beta$ requires the sort of ``mode-dependent renormalization'' that is also required in linear dilaton backgrounds (see e.g. \cite{PP}).

\subsection{Interacting and non-scalar fields}
\label{nonlin}

 Sections \ref{massive} and \ref{massless} above
reviewed boundary conditions at $i^0$ for linear scalar fields.
Parametrizing the space of boundary conditions for an interacting
field theory is more difficult. Unless one imposes the ``fast
fall-off" boundary condition $\alpha=0$, the non-linear interactions
 become strong near infinity and are hard to control.
However, one may linearize the space of boundary conditions about
$\alpha=0$.  Infinitesimal deformations of the boundary conditions
are described by the addition of some linearized solution $\delta
\Phi$ (which is again a linear combination of the modes $\Phi_{q,
\vec j}$ from sections \ref{massive} or \ref{massless}).  In this
way, it is meaningful to vary even a non-linear theory with respect
to $\alpha$, so long as one evaluates all such variations at $\alpha
=0$. A similar structure is commonly used to discuss boundary
conditions of massive scalar fields in AdS (see e.g. \cite{BFS}),
and is the best that one can expect for masses sufficiently far
above the Breitenloher-Freedman bound, where they correspond to
non-renormalizable deformations of the dual field theory.

For the sake of clarity, we have concentrated on scalar field
theory.  The generalization to fields of higher spin is
straightforward.  For concreteness, let us briefly discuss the case
of the gravitational field itself. Linearized gravitational
fluctuations about asymptotically flat space are similar to the
linear scalar solutions reviewed above (see e.g.
\cite{Beig,BS,KSAF}. We may take, e.g., the boundary conditions of
\cite{ABR} (for $d=4$) or \cite{MM} (for $d \ge 4$) to define a
notion of ``fast fall-off.'' Only these boundary condtions are
asymptotically flat. Other boundary conditions which break
asymptotic flatness may then be studied perturbatively, much as was
done for the massless scalar field. This will be sufficient to
construct the asymptotically flat analogue of the `boundary
correlators' used in the AdS/CFT dictionary, which are related to infinitesimal variations of  the path integral with respect to the boundary conditions.  The
issue of finite deformations of asymptotic flatness is more
complicated, however. While a reasonable theory of such deformations
may exist, it is clear from e.g. \cite{BS} in $d=4$ (or \cite{KSAF}
in higher dimensions) that such deformations destroy the entire
asymptotic structure near spatial infinity.   One expects that such
deformations correspond to non-renormalizable deformations of the
dual theory.

\subsection{A Warning about Locality}
\label{warning}

In the above sections we described boundary conditions at spatial infinity in
terms of a function $\alpha$ on ${\cal H}$. This presentation was
chosen to maximize similarity with the asymptotically AdS case,
where boundary conditions are conveniently represented by functions
on the conformal boundary.  However, we warn the reader that the
corresponding notion of locality on ${\cal H}$ is less useful than
on the analogous AdS boundary.

This is not a surprise from the standpoint of gauge/gravity duality.
In AdS/CFT, it is well known that the local properties of the CFT are
related to the asymptotic structure of AdS space.  One sees this
already at the level of symmetries: certain (asymptotic) AdS
isometries induce a dilation on the conformal boundary, so that
taking a bulk operator to the boundary naturally results in a local
dual operator. Similarly, the wave equation associates point sources
on the boundary with a position-dependent length scale in the bulk
which goes to zero at the boundary.  Since it was observed in
\cite{PP,unpub} that such properties fail to hold with either linear
dilaton or asymptotically flat boundary conditions, we may expect
the corresponding locality properties to fail as there well.

Let us take a more precise look at this connection. Since
deformations of AdS boundary conditions correspond to the addition
of CFT sources, the `locality' of such sources should be reflected
in corresponding local properties of the boundary conditions.
Indeed, such a locality property was pointed out in \cite{Witten} in
the Euclidean context:  Consider a deformation of AdS boundary
conditions described by some $\delta \alpha$ of compact support on
the boundary. A given bulk solution $\phi$ will be deformed in a
complicated way, even at parts of the boundary far from the support
of $\delta \alpha$.  However, away from the support of $\delta
\alpha$, the new solution will still respect the {\it original}
boundary conditions.  One often says that $\delta \phi$ is
``normalizable'' outside the support of $\delta \alpha$.  The same
is true in the Lorentzian setting where, as shown in
\cite{HKLL1,HKLL2}, in the Poincar\'e patch one may choose the
deformation to vanish in an open set whose intersection with the
boundary contains all points outside the support of $\delta \alpha$.

In contrast, this feature does {\it not} hold in our asymptotically
flat context.  While $\delta \alpha$ controls the leading $\rho$
behavior, it does not provide the same `local' description of other
non-normalizable terms in $\Phi$.  It is instructive to compute the
`bulk-boundary propagator' $G_\partial$ for asymptotically flat
space:    One begins with a bulk Green's function $G(x,x')$
satisfying, say, Feynmann boundary conditions. One then takes $x^a =
\rho \hat \eta^a$ for some spacelike unit vector $\hat \eta^a$ and
considers the large $\rho$ limit.  To obtain a finite result, one
rescales the limit by the same function of $\rho$ as is used to
define the boundary value $\beta$ in (\ref{ab}).

One finds
\begin{equation}
G(x,x') \sim m^{d-2} \frac{e^{-m|x-x'|}}{(m|x-x'|)^{\frac{d-1}{2}}}
= m^{d-2}  \frac{e^{-m\rho} e^{+m x'{}^a \hat \eta_a}}
{(m\rho)^{\frac{d-1}{2}}} \left(1 + {\cal O}(\rho^{-1})\right),
\end{equation}
so that
\begin{equation}
\label{bb} G_\partial(\hat \eta, x') \sim -\sqrt{\frac{2m}{\pi}}
 \lim_{\rho \rightarrow \infty}    \rho^{\frac{d-1}{2}}   e^{m\rho}   G(x,x') = - m^{\frac{d-2}{2}} \sqrt{  \frac{2 }{\pi }  }   e^{m x'{}^a \hat \eta_a}.
\end{equation}
But now if we consider $x'{}^a = \rho \hat \eta'{}^a$ as $\rho \rightarrow \infty$, we find that (\ref{bb}) diverges whenever $\hat \eta'{}^a \hat \eta_a > 0$.   The non-normalizable behavior is not localized at the point  $\hat \eta \in {\cal H}$.

How is this feature to be interpreted? We argue in the rest of this
work that it is merely another sign of non-locality in the dual
theory. In particular, we are able to calculate boundary two-point
functions in section \ref{concrete}. We also show in section
\ref{LST} that the above feature also arises for the linear dilaton
background dual to little string theory \cite{ABKS}, and is
therefore not an obstruction to the existence of a meaningful dual.

\section{Boundary Operators, Path integrals, and the S-matrix}

\label{formal}

In anti-de Sitter space, boundary correlators are variations of
either the partition function \cite{Witten} (in Euclidean signature)
 or of a transition amplitude $\langle \psi_+  | \psi_-
\rangle$ \cite{GKP,BLK,BLKT} (in Lorentz signature, see
\cite{Lorentz} for details).  In the latter case, the variation is
performed holding $|\psi_- \rangle $ fixed in the far past (retarded
boundary conditions) and holding $| \psi_+ \rangle$ fixed in the far
future (advanced boundary conditions).  By letting $|\psi_+ \rangle$
and $|\psi_- \rangle$ range over a complete set of states, one
defines a full boundary operator. It is such operators which are
most naturally dual to CFT operators under the AdS/CFT
correspondence.  Up to a certain rescaling, they are simply boundary
limits of bulk operators (see e.g. \cite{limits}).

Our goal here is to investigate the analogous construction in
asymptotically flat space. For concreteness we again consider scalar
field theory, but the analysis generalizes directly to higher spin
fields. A study of first variations will motivate a definition of
`boundary operators.' We then address higher variations and find
that issues involving contact terms are more complicated than in AdS
space.  Nevertheless, boundary n-point functions may be defined
directly in terms of the above-mentioned boundary operator. We will
return to these issues again in section \ref{concrete}.

We begin with a (Lorentz-signature) path integral of the form
$\langle \psi_+ | \psi_- \rangle = \int D \phi e^{iS}$, where the
integral is only over fields $\phi$ satisfying the fast fall-off
boundary conditions\footnote{In the massive case.  In the massless
case we assume that, as in section \ref{bcs}, some split of modes
has been made into $\Phi$ and $\tilde \phi$ and that a boundary
condition has been chosen to enforce $\Phi=0$.} of section
\ref{bcs}.  Note that we have absorbed factors containing the
wavefunctions of the states $|\psi_\pm \rangle$ into $S$.  These
factors contribute extra boundary terms to $S$ at the past and
future boundaries $\Sigma_\pm$.   While they are localized on
$\Sigma_\pm$, the particular form of these boundary terms is
generically non-local within $\Sigma_\pm$.

To deform our boundary conditions by an infinitesimal amount
$\alpha$, we shift the domain of integration by some (infinitesimal)
$\Phi$ which satisfies either (\ref{ab}) or (\ref{mlessab}) for this
$\alpha$.  We require $\Phi$ to be a solution to the classical
equations of motion up to some fast fall-off configuration.  That
is, there must be some fast fall-off configuration $\tilde \phi$
(which need not be a solution) such that $\Phi - \tilde \phi$ is a
classical solution\footnote{This condition is to be understood at
leading order in $\hbar$ and may receive quatnum corrections. It is
also interesting to ask what happens if one shifts the domain of
integration by a non-normalizable configuration which as $\hbar
\rightarrow 0$ differs from any solution to the classical equation
of motion by a non-normalizable term.   However, in this case, it is
not clear that the deformed path integral is well-defined.
Certainly, the semi-classical approximation breaks down as there are
no stationary points in the domain of integration.} .

To vary $\langle \psi_+ | \psi_- \rangle $, we must understand how
the action $S$ depends on the boundary conditions.   This depends on
how both the bulk dynamics and the states $|\psi_\pm \rangle$ vary
with $\Phi$.  In AdS, one would choose $\Phi$ to vanish on
$\Sigma_\pm$ so as to preserve advanced boundary conditions on
$|\psi_+ \rangle$ and retarded boundary conditions on $|\psi_-
\rangle$.  However, as discussed in section \ref{warning}, such a
choice is not in general possible in the asymptotically flat
context.  Thus, we must allow an arbitrary deformation of $S$ on
$\Sigma_\pm$ and attempt to define our boundary operators so that
they are independent of this ambiguity. We will, however, require at
each perturbative order in $\Phi$ that i)  $S$ yields the same bulk
equations of motion as for $\Phi=0$ and that ii) $S$ be stationary
on classical solutions.

Such principles do not fully determine the desired extension of $S$,
but they do constrain the possibilities.  Suppose that the action
for $\Phi=0$ takes the standard form
\begin{equation}
S_{0} = - \int_{\cal M} \left( \frac{1}{2}\partial \phi^2 + V( \phi)
\right) \sqrt{-g} + \int_{\Sigma_+ \cup \Sigma_-} S_{\pm}.
\end{equation}
Here ${\cal M}$ represents a volume of spacetime to the future
 of a Cauchy surface $\Sigma_-$  on which $\psi_-$ is specified, and
 to the past of an analogous $\Sigma_+$ on which $\psi_+$ is
 specified.  The boundary terms $S_{\pm}$ on $\Sigma_\pm$
 depend on the details of the states $\psi_\pm$. In terms of
$\tilde \phi = \phi - \Phi$, we seek an action of the form $S =
S_0[\phi] + S_1[\Phi, \tilde \phi]$ where $S_1$ is  a boundary term
linear in $\Phi$ such that variations $\delta S$ performed holding
$\Phi$ fixed vanish on solutions (to first order in $\Phi$). One
finds
\begin{equation}
\label{varyS0} \delta S_{0} =  \int_{\cal M} \sqrt{-g} \left(
\nabla^2\phi - V'(  \phi)\right)  \delta  \phi - \int_{\partial \cal
M}  \sqrt{-h} (n^a\partial_a  \phi) \delta  \phi +
\int_{\Sigma_+ \cup \Sigma_-} B_{\pm}.
\end{equation}
Here $\partial {\cal M}$ denotes only the boundary of $\partial
{\cal M}$ at spatial infinity; i.e., the part of ${\cal H}$
which lies between the Cauchy surfaces $\Sigma_\pm$.

The bulk term in (\ref{varyS0}) contains just the desired equations
of motion.  Since $\delta \phi = \delta \tilde \phi$ and $
\int_{\partial \cal M} \sqrt{-h} (n^a\partial_a  \tilde \phi) \delta
\tilde \phi$ vanishes for any normalizable solution $\tilde \phi$,
on appropriate solutions we find
\begin{equation}
\label{vary2S0} \delta S_{0} = - \int_{\partial \cal M}
\sqrt{-h} (n^a\partial_a \Phi) \delta \tilde  \phi +
\int_{\Sigma_+ \cup \Sigma_-} B_{\pm}.
\end{equation}
The term at $\partial {\cal M}$ in (\ref{vary2S0}) does not vanish
and will be cancelled by the variation $\delta S_1$. One would like
to use the condition $B_\pm =0$ to impose boundary conditions on
solutions at $\Sigma_\pm$, but it is not clear whether such
conditions are compatible with the equations of motion for $\Phi
\neq 0$.  To achieve compatibility, the boundary terms at
$\Sigma_\pm$ may need to be corrected by a term in $S_1$.  Thus,
$S_1$ must be of the form
\begin{equation}
S_1 = \int_{\partial \cal M}  \sqrt{-h}  (\tilde  \phi  + c )
(n^a\partial_a  \Phi)  + \int_{\Sigma_+ \cup \Sigma_-} B_\pm^{(1)} \Phi + O(\alpha^2),
\end{equation}
where $c$ is some fixed function on $\partial {\cal M}$ and
$B_\pm^{(1)}$ are linear operators on $\Sigma_\pm$.  Below we
take $c=0$, but a more general choice merely shifts our boundary
operators by some set of $c$-numbers.  We will, however, need to define our boundary operators in a way that is independent of $B_\pm^{(1)}$.

We are now ready to vary the boundary conditions in our path
integral, shifting the region of integration by $\delta \phi =
\delta \Phi$. With $S=S_0+ S_1$, this variation yields
\begin{eqnarray}
\label{varyamp} \delta \langle \psi_+ | \psi_- \rangle
\Big|_{\Phi=0} &=& \int D\phi \ i e^{iS} \delta S\Big|_{\Phi=0} \cr
&=& \int D\phi \ i e^{iS} \Biggl[  \int_{\cal M} \sqrt{-g} \left(
\nabla^2\phi - V'( \phi)\right)  \delta  \Phi  + \int_{\partial \cal
M}  \sqrt{-h}\ \phi n^a \overleftrightarrow{\partial_a} \delta \Phi
\cr && \hspace*{2.5cm} + \int_{\Sigma_+ \cup \ \Sigma_-}
B_\pm(\delta \Phi)  \Big|_{\Phi=0}  +   \int_{\Sigma_+ \cup \
\Sigma_-} B_\pm^{(1)} \delta \Phi  \Biggr]  \cr &=& i \int_{\partial
\cal M} \sqrt{-h}  \ \langle \psi_+ | \hat  \phi  |\psi_- \rangle
n^a \overleftrightarrow{\partial_a} \delta \Phi  +   \int_{\Sigma_+
\cup \Sigma_-} B_\pm^{(1)} \delta \Phi,
\end{eqnarray}
where in the last step we have used the fact that the matrix
elements of both $B_\pm$ and the equations of motion
vanish\footnote{While less familiar, the vanishing of $\langle
\psi_+ | B_\pm | \psi_- \rangle$ is established in the same manner
that one demonstrates that the vanishing of corresponding matrix
elements of the equations of motion. One shifts the integration
variable by a normalizable configuration and notes that this changes
neither the measure nor the domain of itnegration. Thus, the change
in the path integral is zero, though by computation it is
proportional to a linear combination of the above matrix elements.
By considering all such shifts, one shows that all of these matrix
elements vanish separately.} and we have introduced the local bulk quantum
field operator $\hat \phi(x)$.  We shall reserve the symbol
$\phi(x)$ for c-number field configurations, such as classical
solutions or configurations over which one integrates in the path
integral.

We wish to define a boundary operator which is independent of
$B_\pm^{(1)}$, as this term is associated with the arbitrary
extension of the state $|\psi_\pm \rangle$ to non-zero $\Phi$.  It
is thus natural to define the boundary operator $\hat \phi_\partial
(\alpha)$ to be ($-i$ times) the part of (\ref{varyamp}) given by a
local integral over $\partial {\cal M}$:
\begin{equation}
\label{bop} \hat \phi_\partial(\alpha) :=  \int_{\partial  {\cal M}}
\sqrt{-h} \  \hat \phi (x)  n^a
\overleftrightarrow { \frac{\partial}{\partial {x}^a} }\left(
\Phi[\alpha] \right) (x),
\end{equation}
for any function $\alpha$ on ${\cal H}$.  Here $\Phi[\alpha]$ is any
solution associated with $\alpha(\eta)$ through (\ref{ab}) or
(\ref{mlessab}).   This is a direct analogue of the familiar
structure from AdS space, and in particular parallels the
construction of AdS asymptotic creation and annihilation operators
in \cite{steve}.  For $m^2 > 0$, the discussion at the end of
section \ref{massive} implies that (\ref{bop}) is well-defined when
acting on a dense set of states.

We may also consider the higher correlators
\begin{equation}
\label{bnpt} \langle \psi_+ | \hat \phi_\partial(\alpha_1) \hat
\phi_\partial(\alpha_2) \dots \hat \phi_\partial(\alpha_k) |\psi_-
\rangle.
\end{equation}
These Wightman functions are defined directly by repeated
application of the boundary operators (\ref{bop}) to the state
$|\psi_- \rangle$, though through (\ref{bop}) we see that they satisfy
\begin{eqnarray}
\label{Wkpt} && \langle \psi_+ | \hat \phi_\partial(\alpha_1) \hat
\phi_\partial(\alpha_2) \dots \hat \phi_\partial(\alpha_k) |\psi_-
\rangle = \int_{\partial {\cal M}} \sqrt{-h} d^{d-1} \eta_1   \
\dots \ \int_{\partial {\cal M}} \sqrt{-h} d^{d-1} \eta_k
 \cr &\times& \Phi[\alpha_1](x_1) \dots \Phi[\alpha_k](x_k)   n_1^a
\overleftrightarrow { \frac{\partial}{\partial {x_1}^a} } \dots n_k^a
\overleftrightarrow { \frac{\partial}{\partial {x_k}^a} }  \langle \psi_+ |  \hat \phi(x_1) \ \dots \
\hat \phi (x_k)  |\psi_- \rangle. \ \ \ \ \ \
\end{eqnarray}
Here the $(\rho_i, \eta_i)$ are the hyperbolic coordinates of $x_i$.

It is also interesting to discuss time-ordered boundary correlators defined by
\begin{eqnarray}
\label{Tkpt} && \langle \psi_+ |  T \left( \hat  \phi_\partial
(\alpha_1) \  \dots   \ \hat \phi_\partial (\alpha_n)   \right)
|\psi_- \rangle := \int_{\partial {\cal M}} \sqrt{-h} d^{d-1} \eta_1
\ \dots \ \int_{\partial {\cal M}} \sqrt{-h} d^{d-1} \eta_k
 \cr   &\times& \Phi[\alpha_1](x_1) \dots \Phi[\alpha_k](x_k)   n_1^a
\overleftrightarrow { \frac{\partial}{\partial {x_1}^a} } \dots n_k^a
\overleftrightarrow { \frac{\partial}{\partial {x_k}^a} }  \langle \psi_+ | T \left( \hat \phi(x_1) \ \dots \
\hat \phi (x_k) \right)  |\psi_- \rangle. \ \ \ \ \ \
\end{eqnarray}

In the AdS context, time-ordered $k$-point boundary correlators are
$k$th functional derivatives of the transition amplitude $\langle
\psi_+ | \psi_- \rangle$.   However, due to contact terms, this
relation holds only when the supports of the variations do not
overlap. With AdS asymptotics, it is not difficult to choose the
$\Phi[\alpha_i]$ to have non-overlapping support in the bulk
spacetime.  However, this is {\it not} in general possible in
asymptotically flat space.  As noted in section \ref{warning}, even
for functions $\alpha_1, \alpha_2$ with well separated supports on
${\cal H}$, the supports of the bulk functions $\Phi[\alpha_1],
\Phi[\alpha_2]$ must overlap.  Thus, for $k>1$, time-ordered
$k$-point functions are variations of the path integral only up to
i) terms at $\Sigma_\pm$ as in the discussion of one-point functions
and ii) additional (typically divergent) terms associated with
contact terms in the bulk. We refer to terms of type (ii) as
``contact terms'' even though they occur for boundary operators with
disjoint supports.

\section{Boundary Correlators}
\label{concrete}

Having discussed the general structure of our boundary operators, we
now compute boundary two-point functions. We attempt two
computations, though only one succeeds. The first (section
\ref{T2pt}) is an attempt to follow the analogue of the procedure
\cite{GKP} used by Gubser, Klebanov, and Polyakov for AdS/CFT.
Unfortunately, this approach suffers from the divergent non-local
contact terms mentioned in section \ref{formal} above.  As a result,
it is unclear which non-local terms one should substract to obtain
the correct (finite) two-point function.  The same non-local contact
terms prevent a straightforward application of the method of
\cite{Witten}.

On the other hand, we show (section \ref{Wmethod}) that the boundary
{\it Wightman} two-point functions are readily calculated using the
basic definition (\ref{bop}).  The result is finite and unambiguous,
and it leads to a (well-defined) time-ordered boundary two-point
function with the expected analytic structure.  To gain additional
perspective on the above issues, we consider the same calculations
in linear dilaton backgrounds in section \ref{LST} and find similar
results.

\subsection{Boundary 2-point functions via variations of the action}
\label{T2pt}

We noted in section \ref{formal} that our time-ordered boundary two-point
functions are the second variations of a bulk partition function, up
to contact terms and terms at $\Sigma_\pm$. In the limit in which
the bulk system is semi-classical, this variation is just the
on-shell variation of the bulk action:

\begin{equation}
\label{2act} \frac{ \langle |  T\left( \hat \phi_\partial (\delta_1
\alpha) \  \hat \phi_\partial (\delta_2 \alpha) \right) |0 \rangle}
{\langle 0|0 \rangle} \approx  - \delta_1 \delta_2 S + \ {\rm
contact \ terms} \ + \ {\rm terms \  at} \ \Sigma_\pm.
\end{equation}
To compute such correlators, we study variations of the
semi-classical action and attempt to remove the extraneous terms.
This is essentially the approach to calculating boundary correlators
(in AdS) advocated in \cite{GKP}.  Since the goal is to obtain
time-ordered vacuum correlators, one expects that one may avoid
consideration of future and past boundary terms by analytic
continuation to Euclidean signature and taking $\Sigma_\pm$ to
infinity.  We shall do so below.

The variation $\delta_1 \delta_2 S$ evaluated at $\phi=0$ depends
only on the part of the action quadratic in $\phi$; it is
independent of any couplings of $\phi$ to itself or to any other
fields (including gravity).    The calculation is similar to
(\ref{varyS0}) and yields:
\begin{equation}
\label{v2} \delta_1 \delta_2 S = - \int_{\partial \cal M} \sqrt{-h} \ \delta_1
\Phi[\alpha] \ n^a \partial_a \delta_2 \Phi [\alpha],
\end{equation}
where $\delta_1 \alpha$, $\delta_2 \alpha$ are now functions on $S^{d-1}$ and  $\delta_1 \Phi[\alpha], \delta_2 \Phi[\alpha]$ are solutions (up to normalizable terms) associated with $\delta_1 \alpha$, $\delta_2 \alpha$ through the Euclidean version of (\ref{ab}) or (\ref{mlessab}).

As usual, the calculation is most straightforward using normal modes.  Thus we take
\begin{equation}
\label{a1a2}
\delta_1 \alpha = \epsilon_1 Y_{\vec \ell_1}, \ \ \
\delta_2 \alpha = \epsilon_2 Y_{\vec \ell_2},
\end{equation}
where $Y_{\vec \ell_2}$ are harmonics on $S^{d-1}$ and $\epsilon_1,
\epsilon_2$ are (infinitesimal) constants.  The notation here
matches our previous notation for harmonics on the sphere; e.g.
$|\vec \ell |^2 = \ell (\ell +d -2)$. For a massive field $\phi$ we
have
\begin{eqnarray}
\label{Euclmodes} \delta_1 \Phi [\alpha] =    \epsilon_1
\rho^{\frac{2-d}{2}} I_{\nu_1} (m\rho) Y_{\vec \ell_1}, \cr \delta_2
\Phi [\alpha] =  \epsilon_2 \rho^{\frac{2-d}{2}} I_{\nu_2} (m\rho)
Y_{\vec \ell_2},
\end{eqnarray}
where $\nu = \ell + \frac{d-2}{2}$.

To evaluate (\ref{v2}), we make use of the asymptotic expansion:
\begin{equation}
\label{Inu} I_{\nu} (z) = \frac{1}{\sqrt{2\pi z}}A_\nu (z) + \frac{
\cos \left( (\nu + \frac{1}{2}) \pi  \right) }{\sqrt{2\pi z}}
B_\nu(z),
\end{equation}
where
\begin{eqnarray}
\label{AB}
 A_\nu (z) &:=&  e^z \sum_{k=0}^\infty \frac{(-1)^k}{(2z)^k}
\frac{\Gamma(\nu + k + \frac{1}{2}) } {k! \Gamma(\nu - k +
\frac{1}{2}) }, \ \ \ {\rm and}  \cr
 B_\nu (z) &:=& e^{-z} \sum_{k=0}^\infty
\frac{1}{(2z)^k} \frac{\Gamma(\nu + k + \frac{1}{2}) }{k! \Gamma(\nu
- k + \frac{1}{2}) }.
\end{eqnarray}
We have obtained (\ref{Inu},\ref{AB}) from \cite{GR}, though in
\cite{GR}, the coefficient of $B_\nu$ in (\ref{Inu}) involves
$\exp(\pm (\nu + \frac{1}{2}) \pi i)$ instead of the cosine.  This
ambiguity is described as a ``Stokes Phenomenon.'' Since we require
a real solution, we have simply taken the average of the two
expansions. However, the result below is identical if one uses
$\exp(\pm (\nu + \frac{1}{2}) \pi i)$ and chooses the sign in a way
which depends only on $\nu$.

Inserting expressions (\ref{Inu},\ref{AB}) into (\ref{v2}) yields
three kinds of terms:

\begin{enumerate}[i)]

\item Terms involving $B_{\mu_1} \partial_\rho B_{\mu_2}$,
which are too small to contribute as $\rho \rightarrow \infty$.

\item Terms involving $A_{\mu_1}  \partial_\rho B_{\mu_2}$ and $B_{\mu_1} \partial_\rho A_{\mu_2}$.
These give finite contributions, which cancel against each other.

\item \label{AA} Divergent terms involving the expressions
$A_{\mu_1} \partial_\rho A_{\mu_2}$.
\end{enumerate}
Recall that the form of the action has only been fixed up to ${\cal
O}(\alpha^2)$ terms.  It is natural to choose such terms to
precisely cancel the terms of type (\ref{AA}), which arise only from
non-normalizable terms in the expansion of (\ref{Inu}).  As we will
discuss in section \ref{LST} below, this appears to be the analogue
of the procedure followed in \cite{Shiraz} for massless fields in a
linear dilaton background.  One may hope that this is equivalent to
subtracting the extra ``contact terms'' noted above, together with
any other contact divergences inherent in the correlator itself.
After making this choice, we find
\begin{equation}
\label{T2result} \langle 0| T\left( \hat \phi_\partial(Y_{\vec
\ell_1}) \hat \phi_\partial(Y_{\vec \ell_2}) \right) | 0 \rangle = 0
\end{equation}
for all $m^2 > 0$.

We may also compute the correlator for $m=0$. In this case, the
radial mode functions are simply $\rho^\lambda$ for appropriate
$\lambda$; there is no apparent mixing between normalizable modes
$(\tilde \phi)$ and non-normalizable modes $(\Phi)$. Thus,
subtracting the $m=0$ analogue of the type (\ref{AA}) terms again
yields (\ref{T2result}).

The fact that (\ref{T2result}) vanishes identically suggests caution
in interpreting the result.  Indeed, as remarked above, the
particular subtractions we have used are far from well-justified.
Recall that for $m^2 > 0$ our subtractions are non-analytic in
$q^2$, and thus do not qulitatively differ from the sort of finite
remainder terms that one would expect.  On general grounds one might
also expect to require a non-analytic subtraction for $m^2=0$
(though we did not use one there).  Thus, this approach to
calculating the 2-point function appears to be inherently ambiguous.

We note that the method of \cite{Witten} will meet with similar
problems: In AdS/CFT it avoids divergences by working with separated
operators, but our asymptotically flat computations generate
non-local ``contact terms'' which can diverge even at separated
points. Thus, we must seek another approach.

\subsection{Boundary correlators via the Wightman Function}
\label{Wmethod}

One might like to calculate 2-point functions directly from the
basic definition of the boundary operator (\ref{bop}).  This is
indeed possible if one focusses on Wightman functions, and such an
approach turns out to have several advantages.  For example, in any
local field theory, the Wightman function is a well-defined
bi-distribution, meaning that it is finite when integrated against
two smooth functions which behave appropriately at infinity.  There
are thus no divergences from contact terms. The same is true of
Wightman boundary correlators in AdS, and we will see that the same
is again true for asymptotically flat boundary two-point functions.

The computation is straightforward using the representation of the
bulk Wightman function $W(x,x')$ as a sum over
Klein-Gordon-normalized positive frequency modes. Since modes with
$q
> 0$ are positive frequency while modes with $q < 0$ are negative
frequency, and since propagating modes satisfy (\ref{qbound}), for
any $m \ge 0$ we have
\begin{equation} W(x,x') =\int_{q > \frac{d-2}{2}} dq  \sum_{\vec j}  \frac{2 }{ \mu |\Gamma(i \mu)|^2}  \phi^*_{q, \vec j}(x)  \phi_{q, \vec
j}(x').
\end{equation}
Here we have used expression (\ref{mnorms}) or (\ref{mlessnorms})
for the Klein-Gordon norms of our modes.

Taking the boundary product of $W$ with two modes $\Phi_{q_1, \vec
j_1}, \Phi^*_{q_2, \vec j_2}$ with $q \ge \frac{d-2}{2}$ and using
the orthonormality of the $Y_{q, \vec j}$ we find
\begin{eqnarray}
\label{bWr}
 W_\partial(q_1,\vec j_1, q_2, \vec j_2) &: =& \langle 0|
\hat \phi_{\partial}(Y_{q_1, \vec j_1}) \hat
\phi_{\partial}(Y^*_{q_2, \vec j_1}) |0 \rangle \cr &=& \frac{2 }{
\mu |\Gamma(i \mu)|^2} \delta(q_1 - q_2) \delta_{\vec j_1,\vec j_2}
\ \ \ {\rm for} \ \ q_1,q_2 \ge \frac{q-2}{2}.
\end{eqnarray}
The boundary Wightman function vanishes for other modes. The result
is finite, non-zero, independent of $m$, and insensitive to the
Bessel Stokes phenomenon.

From the result (\ref{bWr}), one may now unambiguously compute the
associated time-ordered 2-point function.  To do so, one need only
write (\ref{bWr}) as
 \begin{equation}
 W_\partial(q_1,\vec j_1, q_2, \vec j_2) : =
 \int_{0}^{\infty} \frac{d \mu}{\pi  |\Gamma(i \mu)|^2}
 W^{\cal H}_{q(\mu)}(q_1,\vec j_1, q_2, \vec j_2),
 \end{equation}
 where $q(\mu) = \sqrt{\mu^2 + \left( \frac{d-2}{2}
 \right)^2}$ and
 \begin{equation}
 W^{\cal H}_{q}(q_1,\vec j_1, q_2, \vec j_2) = \frac{\pi}{q} \delta(q-q_1)
 \delta(q-q_2)\delta_{\vec j_1,\vec j_2}
 \end{equation}
is the Wightman function on ${\cal H}$ for a free scalar field of
mass $q$. The is essentially the K\"allen-Lehman representation of
the Wightman function on ${\cal H}$. The time-ordered two-point
function is then
 \begin{equation}
 \langle 0| T\left( \hat \phi_\partial(Y_{\vec
\ell_1}) \hat \phi_\partial(Y_{\vec \ell_2}) \right) | 0 \rangle =
\int _{0}^{\infty} \frac{d \mu}{\pi  |\Gamma(i \mu)|^2}
 G^{F,{\cal H}}_{q(\mu)}(q_1,\vec j_1, q_2, \vec j_2),
 \end{equation}
where $G^{F,{\cal H}}_{q(\mu)}$ is the Feynmann Green's function on
${\cal H}$ for a free scalar field of mass $q$.

Note that there is a branch cut beginning at $q=\frac{d-2}{2}$
associated with the continuum of propagating bulk states.  Indeed,
one sees that the analytic structure is determined by the
K\"allen-Lehman spectral function of the boundary operator
$\phi_\partial$, which is in turn determined directly  by the
spectrum of bulk states.

\subsection{Little String Theory}

\label{LST}

Despite our success with methods based on Wightman functions, we
have seen that computations of asymptotically flat boundary 2-point
functions by analogy with either \cite{GKP} or \cite{Witten} faced
serious difficulties.  Now, as mentioned in the introduction, linear
dilaton backgrounds dual to little string theory share many features
of asymptotically flat spacetimes.  As a result, one may hope to
gain further insight by pursuing this analogy in detail. To that
end, we now consider boundary two-point functions associated with a
scalar field in the linear dilaton background dual to little string
theory (namely, in the near-horizon solution for $N$ coincident
$NS5$-branes \cite{ABKS}).  We will see that, despite the evident
success of the Gubser-Klebanov-Polyakov method in this context
\cite{Shiraz,NR}, the same issues identified in section \ref{T2pt}
also arise in linear dilaton backgrounds.


Let us first briefly review the linear dilaton spacetimes of
interest.  In the string frame, the near horizon description of $N$
coincident $NS5$-branes takes the familiar form
\begin{eqnarray}
ds^2_{string} = dx_6^2 + dz^2 + \frac{N}{m_s^2}d \Omega_3^2, \cr
g = g_0 e^{-\frac{z m_s}{\sqrt{N}}},
\end{eqnarray}
where $dx_6^2$ is the 6-dimensional Euclidean metric and $m_s$ is
the string mass scale. In the strong coupling regime at large
negative $z$, the physics is more properly described by the
near-horizon metric of $N$ $M5$-branes on an $S^1$. However, we will
be interested only in the asymptotics at large $z$ where the
corrections are heavily suppressed.

It is natural to consider a scalar field minimally coupled to the
Einstein frame metric.  This metric takes the tantalizing form
\begin{equation}
\label{LDE} ds^2 = \rho^2 (dy_6^2  + \frac{1}{16} d\Omega^2_3) + d
\rho^2,
\end{equation}
where $\rho = \frac{4 \sqrt{N}}{m_s} e^{zm_s/4 \sqrt{N}}$ and
$dy_6^2$ is again the 6-dimensional Euclidean metric, but with
rescaled coordinates  $y_i = \frac{m_s}{4\sqrt{N}}x_i$.   While
(\ref{LDE}) is not asymptotically flat, the metric components involve the same
powers of $\rho$ as in flat Minkowski space\footnote{From this point
of view, the $S^3$ in (\ref{LDE}) is on the same footing as the
$\mathbb{R}^6$.  Yet only the $\mathbb{R}^6$ forms the spacetime of
little string theory; the $S^3$ is associated with an internal
symmetry.  It is possible that the fate of the asymptotically flat
${\cal H}$ is more similar to this $S^3$ than to the $\mathbb{R}^6$.
However, given that we expect a non-local theory, this distinction
may not be crucial at this level.}.

Massless scalar fields minimally coupled to (\ref{LDE}) were studied
in \cite{Shiraz}.  We now consider massive minimally coupled
fields\footnote{It is not clear that string theory contains such
fields, but it does contain close analogues.  For example,
$D0$-branes naturally couple to a metric that can be written in the
form (\ref{LDE}), but with a different coefficient in front of $
d\Omega^2_3$ and a further rescaling of the $\mathbb{R}^6$.}.
Solutions to the massive scalar wave equation are given by
\begin{eqnarray}
 \Phi_{k, \vec j} &=& \frac{1}{(2\pi)^3} Y_{\vec j} e^{ik_ay^a}
I_\nu(m \rho), \cr
 \Phi_{k, \vec j} &=& \frac{1}{(2\pi)^3}  Y_{\vec
j} e^{ik_ay^a} K_\nu(m \rho),
\end{eqnarray}
with $\nu = \sqrt{k^2 + 16 j(j+2) + 16}$ and $\vec j$ labeling a
complete set of states in $SO(4)$. Since the string coupling goes to
zero at large $\rho$, M-theoretic corrections to (\ref{LDE}) die off
faster than $e^{-\lambda \rho}$ for any $\lambda$ and the
corrections to these mode functions will be correspondingly small,
though they might result in some ``mixing'' through the addition of
a further normalizeable piece  $C K_\nu(m\rho)$ to the
``non-normalizable'' mode $I_\nu(m \rho)$ associated with some
particular boundary conditions.  Note, however, that $C$ must be
determined by $k^2$ and $j$ ($C = C(k^2, j)$) and that $C$ is real
(at least in Euclidean signature).

Let us now consider the 2nd on-shell variation of the classical
action. As in section \ref{T2pt}, three kinds of terms will be
generated. The finite (type ii) terms again cancel\footnote{For $m^2
> 0$.  It is interesting that this does not occur \cite{Shiraz} for $m=0$.}
for any $C(k^2,j)$,  so long as the Stokes phenomenon is again dealt
with as in section (\ref{T2pt}).  Terms containing $A_{\nu_1}
\partial_\rho A_{\nu_2}$  are divergent.  As in the asymptotically flat case, subtracting such terms yields a time-ordered two-point function which vanishes for any $m^2 > 0$.

However, for the same reasons as in section \ref{T2pt}, the divergences associated with
$A_{\nu_1} \partial_\rho A_{\nu_2}$ terms are again non-analytic in $k$; they do not
have the form of familiar contact terms.
To interpret these divergences, recall that little string theory is
non-local \cite{NSNS}.  We might therefore expect that any
time-ordering operation is more complicated than for a local theory,
and may lead to divergences even at separated points.  In fact, the
same argument as in section \ref{formal} suggests that the variation
of the path integral will differ from the time-ordered two-point
function by non-local contact terms. The key point is that the
linear dilaton background's bulk-boundary Green's function is
non-local in precisely the sense described in section \ref{warning}
for asymptotically flat space. To see this, consider the
non-normalizable solution
\begin{equation}
\Phi = \frac{1}{(2\pi)^6} \int dk \ \Phi_{k, \vec \ell}.
\end{equation}
The leading behavior is $\Phi \sim \delta(x) Y_{\vec \ell} \
\frac{e^{-m \rho}}{\sqrt 2 \pi m \rho}$, but the non-analytic
dependence of $A_\nu$ on $k$ means that $\Phi$ contains subleading
non-normalizable terms not localized at $x=0$.  While this
observation again encourages the subtraction of such divergences, it
raises the disturbing prospect that the remaining finite part may be
contaminated with unwanted (but finite) non-local ``contact'' terms.

Let us also briefly compare the computations for massless fields.
The linear dilaton calculation was performed in \cite{Shiraz}, and
the asymptotically flat case was discussed in section \ref{T2pt}
above. Both calculations obtain a finite answer by subtracting only
divergences analytic in $k$. However, both manifest signs of
non-locality via the need for `mode-dependent renormalization'
\cite{PP,AB,Shiraz}. One apparent difference is that the asymptotically flat
result vanished identically. However, despite the removal only of
divergences analytic in $k$, there is no local bulk-boundary
propagator and it is possible that the finite remainders are again
contaminated by non-local contact terms, and any comparison of the results should proceed with caution.

We see that there is a strong similarity between the issues that
arise in the asymptotically flat and linear dilaton contexts. Of
course, there is also a significant difference.  Namely, as shown in
\cite{Shiraz,NR}, at least for massless scalar fields, one {\it
does} appear to obtain a physically interesting time-ordered
two-point function by applying the analogue of the
Gubser-Klebanov-Polyakov method \cite{GKP} along with a naive
subtraction of leading divergences.    In particular, \cite{Shiraz}
showed that at small momenta such computations precisely agree the
results of the theory on M5-branes.  Furthermore, in the thermally
excited context, \cite{NR} found such computations to precisely
agree with those of \cite{GK} performed in little string theory.
Finally, \cite{LSZLST} argued that the analytic structure of the
two-point function so-obtained properly matches expectations from
the bulk spectrum of states.  The evident success of such
comparisons suggests that a unique prescription for subtracting
non-local contact terms can be established for linear dilaton
backgrounds and thus, perhaps, for asymptotically flat spacetimes as
well.  Unfortunately, for the moment such a prescription remains a
mystery.

In the asymptotically flat context, we saw that computations of
boundary Wightman functions are free of the subtleties discussed
above.  As a result, this method appeared to be preferred over
working with the on-shell action.  In the abstract, the same
argument may be made for linear dilaton backgrounds: the Wightman
calculation is straightforward, and again free of ambiguities. Yet
one may ask if the linear dilaton Wightman calculation can reproduce
the above linear dilaton successes of the on-shell action
method\footnote{It is clear from \cite{Witten} that the two methods
agree for AdS/CFT.}. While we leave a detailed analysis of this
question for future work, we note that the analytic structure of the
time-ordered correlator obtained via our Wightman-function method is
directly tied to the bulk spectrum of states for the same reasons as
in the asymptotically flat case.  In particular, from the above
results we find for any $m^2 \ge 0$ that, up to M-theory
corrections, the boundary Wightman function is
\begin{equation}
\langle 0| \hat \phi_{\partial}(e^{-ik^1_a y^a} Y^*_{\vec \ell_1} )
\hat \phi_{\partial}(e^{ik^2_a y^a} Y_{\vec \ell_2} ) |0 \rangle =
\frac{2}{\mu |\Gamma(i\mu)|^2} \delta^{(6)}(k^1-k^2) \delta_{\vec
\ell_1,\vec \ell_2} ,
\end{equation}
when $\mu = \sqrt{-k^2 - 16j(j+2)-16}$ is real and $k_0>0$, and the
boundary Wightman function vanishes when $\mu$ is imaginary or $k_0
<0$.  The result is finite, and independent of $m$.  As in the
asymptotically flat case, the associated time-ordered correlator is
readily computed using a spectral representation, which is in turn
determined directly by the spectrum of states in the bulk.

\section{Discussion}
\label{disc}

We have proposed a framework in which an AdS/CFT-like correspondence
may be explored for asymptotically flat spacetimes. Deformations of
asymptotically flat boundary conditions are naturally associated
with a Lorentz-signature hyperboloid ${\cal H}$ at spacelike
infinity.  This ${\cal H}$ is the home of our holographic dual, and
we have stressed the analogy with the $\mathbb{R}^{5+1}$ home of
little string theory which lies at spacelike infinity in a linear
dilaton spacetime.

As in AdS/CFT, the basic objects in our correspondence are boundary
correlators, which are related to variations of boundary conditions
at ${\cal H}$.  In contrast, one often considers the S-matrix to be
the fundamental observable in asymptotically flat spacetimes.  It is
natural to expect that these observables are related.  Indeed,
recall that in AdS the time-ordered boundary correlators are related
to on-shell truncated Green's functions and thus to the S-matrix
\cite{steve}. At the formal level, such arguments carry over
directly to our asymptotically flat setting and suggest that
time-ordered boundary Green's functions recode the same information
found in the S-matrix.  In particular, since the bulk-boundary
propagator produces solutions which behave like $e^{m x^a k_a}$ for
real $k_a$ (section \ref{warning}), it is natural to regard our
boundary correlators as an analytic continuation of the S-matrix to
spacelike momenta.  One would like to find a precise form of this
statement through an appropriate treatment of the contact terms and
(IR) divergences from section \ref{T2pt}.  Results for linear
dilaton backgrounds suggest that this is possible, but the details
remain unclear.

Before turning to more  technical issues, we should address a
conceptual concern.  In AdS/CFT, one often describes boundary
operators as inserting particles into the bulk. It is also common to
consider signals which enter through the boundary, propagate
causally through the bulk, and then return to the boundary.  Clearly
no such discussions are possible for a dual theory at spacelike
infinity, since there are no causal curves connecting this boundary
to the bulk.

\FIGURE{\includegraphics[width=1in]{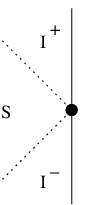}\caption{A
simplified conformal diagram of the AdS boundary.  }
\label{acausal}} However, we wish to emphasize that while such
causal discussions are possible in the AdS context, {\it acausal}
connections between the bulk and boundary are nevertheless central
to the duality.  This is most immediately evident from the
expectation that the boundary theory encodes information inside
large stable black holes, though it can also be seen by considering
horizon-free geometries.  The point is that the CFT must encode the
full bulk dynamics at {\it each} time; i.e., on any Cauchy surface
$C$ in the $S^n \times \mathbb{R}$ spacetime in which it resides.
Thus, on any such $C$ one expects the CFT to holographically encode
even information about the part of the AdS bulk to which it is {\it
not} causally connected. This feature is illiustrated in figure
\ref{acausal}, which provides a simplified conformal diagram of the
AdS boundary.  The black dot represents a Cauchy surface in the
boundary manifold (solid line). The dual theory on this surface must
encode not only bulk data from the regions $I^\pm$ to which it is
causally connected, but also from the causally disconnected region
$S$.

Our main technical results are the computation of time-ordered and
Wightman boundary two-point functions in asymptotically flat
spacetimes.    These calculations raise a number of interesting
issues:

\begin{enumerate}

\item [{$\mathbf m^2 >0$}]  At first glance, the asymptotic behavior
of massive fields near $i^0$ seems to be in direct parallel to the
near-boundary behavior of scalar fields in asymptotically AdS
spacetimes.  The leading asymptotics are a fixed function of $\rho$
times an arbitrary function of the coordinates $\eta$ on $i^0$. In
particular, there is no issue of `mode-dependent renormalization.'

However, attempts to calculate time-ordered two-point functions from
the on-shell action (section \ref{T2pt}) required the cancellation
of divergences non-analytic in the boundary momenta. Such
divergences cannot correspond to local contact terms on the
boundary.  Furthermore, we noted in section \ref{LST} that the same
phenomenon occurs in linear dilaton backgrounds (where a
gauge-gravity duality is well-established \cite{ABKS}).  We
therefore propose that these divergences have the same logical
status as do local contact terms in a local field theory, and that
their non-local nature merely reflects the non-locality of the dual
theory.

Two further forms of evidence were presented in favor of this
viewpoint: First, it was noted that even when considering boundary
operators with disjoint support, in both asymptotically flat and
linear dilaton spacetimes the corresponding bulk calculations {\it
do} involve contact terms.  This is in sharp contrast to the
asymptotically AdS case.  Second, we considered the boundary
Wightman functions for both asymptotically flat and linear dilaton
spacetimes. If the boundary operators are well-defined, then such
boundary Wightman functions {\it must} be finite and should be
calculable directly from the basic definition (\ref{bop}) of the
boundary operators; no subtraction of divergences is allowed.
Indeed, our Wightman functions were both finite and unambiguous, and
they led to similarly well-defined time-ordered correlators.  This
supports the conjecture that meaningful dual theories exist, and
that non-local divergences in the on-shell action merely reflect
complications of non-local theories.

\item [{$\mathbf m^2 =0$}] Massless fields in either asymptotically flat
or linear dilaton backgrounds behave differently from massive
fields.    The dual boundary operators require `mode-dependent
renormalization' and, as a result, there is no canonical map between
boundary conditions and functions on the boundary; i.e., there is no
canonical transformation of our momentum-space boundary correlators
to position space.  Nevertheless, using a natural normalization of
the boundary condition mode-functions, only local divergences appear
when calculating time-ordered boundary two-point functions from the
on-shell action.

On the other hand, as in the massive case, an argument based on
variations of the path integral suggests that non-local contact
terms can in fact arise.    Thus, it is again unclear whether
computations of time-ordered correlators from the on-shell action
can be fully trusted; the finite results may be polluted by
(non-local) contact terms. Such pollution may account for the rather
surprising fact that this method (together with a naive subtraction
of divergences) led to an asymptotically flat boundary two-point
function which vanished identically. In contrast, a computation of
boundary correlators via the bulk Wightmann function was
well-defined.
\end{enumerate}

In the abstract, calculations via the on-shell action in linear
dilaton backgrounds appear (section \ref{LST}) to suffer from the
same difficulties as in asymptotically flat spacetimes.  Yet, there
it is known \cite{Shiraz,NR} that a naive subtraction of divergences
leads in the end to physically useful boundary correlators.  This
success remains mysterious at present but, if it could be
understood, it may indicate how much procedures may be properly
applied to asymptotically flat boundary correlators as well.

Let us now ask how one might move beyond the bulk gravity
approximation.  If a precise relation between our boundary
correlators and the S-matrix were found, it would allow computation
of boundary correlators using known techniques in string
perturbation theory. However, constructing the boundary Wightman
functions directly from string theory might offer a way around the
troublesome divergences. While at present no such technology is
available, the fact that these correlators live on the boundary (and
so are fully gauge invariant) and are defined using `on-shell'
boundary conditions near $i^0$ encourage the belief that they are
well-defined in string theory, and that this problem is merely
technical.

Perhaps the most important remaining issue concerns the role of
symmetries in our proposed duality. It is clear that the bulk
Lorentz group acts on the boundary theory as the corresponding group
of isometries on the hyperboloid ${\cal H}$. A boundary stress
tensor whose integrals give associated conserved quantities was
described in \cite{MM,MMV}.  However, the status of bulk
translations is less clear\footnote{As noted in section
\ref{nonlin}, we choose our boundary conditions on the metric
following \cite{ABR,MM}.  As a result, supertranslations do not act
as symmetries.}.  Because translation killing fields are smaller at
infinity by a factor of $1/\rho$ as compared with rotations and
boosts, the translations naturally leave points of ${\cal H}$
invariant.  However, they need not act trivially on the boundary
theory. Let us label points on ${\cal H}$ with unit vectors $\hat
\rho^a$ in $\mathbb{R}^d$.  In Euclidean signature, under a
translation $x^a \rightarrow x^a + \lambda^a$ we have $\rho
\rightarrow \rho + \lambda^a \hat \rho_a$ for large $\rho$.  For
$m^2 > 0$, the result is that each `boundary condition' $\Phi_{q,
\vec j}$ is multiplied by $\exp(m \lambda^a \hat \rho_a)$.  Since
the bulk vacuum correlators are translation invariant, the
position-space boundary correlators are multiplied by one factor of
$\exp(m \lambda^a \hat \rho_a)$ for each argument.  In order for
this to be a symmetry, the boundary correlators must be invariant;
i.e., they must vanish unless the arguments satisfy $\sum_i \hat
\rho^a_i =0$.

However, our Wightman functions do not appear to satisfy this
condition. This may be related to our lack of success with Euclidean
methods in section \ref{T2pt}.  In particular, the action of
translations on boundary conditions is rather less clear in Lorentz
signature, where the quantity $\lambda^a \hat \rho_a$ grows
arbitrarily large on each ${\cal H}_\rho$ and no expansion in
$\frac{\lambda^a \hat \rho_a}{\rho}$ is uniformly valid. Identifying
the action of translations on the $\Phi_{q, \vec j}$ will therefore
require more sophisticated techniques.  We leave this important
issue for future investigation.

As noted above, our hyperbolic representation of infinity is well
adapted to the Lorentz group.  However, one can imagine other
representations of $i^0$.  For example, it is often useful to
represent $i^0$ by a cylinder which extends in the time direction.
This construction is natural in thermal contexts where one wishes to
work in Euclidean space with periodic time. However, a notable
feature of cylindrical representations is that the metric along the
cylinder (i.e., the `time' direction) does not grow as one
approaches $i^0$.  Thus to some extent the cylinder $\mathbb{R}
\times S^{d-2}$ is merely an infinitesimal region of the hyperboloid
${\cal H}$ near its intersection with the $t=0$ plane. Nonetheless,
it would be interesting to explore this construction in more detail.

The main message of our work is to note that, in asymptotically flat
and linear dilaton spacetimes, the null part of the boundary may not
play a significant role in any gauge/gravity duality.  In contrast,
the dual theory is naturally associated with a part of inifinity
which lies at the endpoints of {\it spacelike} geodesics, and which
is associated with the specification of boundary conditions for the
bulk.  Plane wave spacetimes would also be interesting to analyze
from this point of view.  The boundaries so far understood
\cite{BN,MR} for such spacetimes are null, and thus do not provide
natural homes for dual theories. One would like to explore the
possibility that the identification of an appropriate spacelike
infinity (see e.g. \cite{MR2}) could lead to a self-contained theory dual to plane wave
spacetimes.  Such a result would further clarify the BMN limit
\cite{BMN} of the AdS/CFT correspondence.

\subsection*{Acknowledgments}
The author would like to thank Ofer Aharony, David Berenstein,  Jan
de Boer, Steve Giddings, David Gross, David Lowe, Jim Hartle, Gary
Horowitz, David Lowe, Mark Srednicki, Djordje Minic, Jan Troost,
Mukund Rangamani, and Amitabh Virmani for a number of useful
conversations. This work was supported in part by NSF grant
PHY0354978 and by funds from the University of California.

\appendix

\section{Klein-Gordon Normalizations}

This  appendix computes the Klein-Gordon inner product of modes of
the form (\ref{mmodes}):
\begin{equation}
\label{KGdefA} \langle \tilde \phi^*_{q_1, \vec j_1} ,  \tilde
\phi_{q_2, \vec j_2} \rangle_{KG} := i \int_\Sigma \sqrt{g_\Sigma} \
\phi^*_{q_1, \vec j_1} n^a \overleftrightarrow \partial_a  \tilde
\phi_{q_2, \vec j_2},
\end{equation}
where $\Sigma$ is a Cauchy surface in ${\cal M}$ and $*$ denotes
complex conjugation. Note that (\ref{KGdefA})  is the product of an
$L^2$ inner product of the radial functions
\begin{equation}
\label{L2} \int_{\rho > 0} \frac{d \rho}{\rho} \ K^*_{i \mu_1}(m
\rho) K_{i \mu_2}(m \rho),
\end{equation}
with a Klein-Gordon inner product on ${\cal H}$:
\begin{equation}
\label{KGHdef}
\langle  Y^*_{q_1, \vec j_1} ,  Y_{q_2, \vec j_2} \rangle_{KG, {\cal
H}} := i \int_C \sqrt{\omega_C} \ Y^*_{q_1, \vec j_1} \tilde n^a
\overleftrightarrow \partial_a  Y_{q_2, \vec j_2}.
\end{equation}
Here $C$ is a Cauchy surface in ${\cal H}$ with induced volume
element $\sqrt{\omega_C}$ and future-pointing unit (in ${\cal H}$)
normal $\tilde n^a$. In (\ref{L2}), $\mu_i = \sqrt{q^2 -
\frac{d-2}{2}}$.

The $L^2(\mathbb{R}^+, \frac{d \rho}{\rho})$ factor may be calculated by realizing that $K_{i\mu}(e^\lambda)$ are eigenfunctions of the operator $- \partial_\lambda^2 + m^2 e^{2 \lambda}$ in $L^2({\mathbb R}, d \lambda)$ with eigenvalue $\mu^2$; i.e., by mapping the calculation to a familiar scattering problem in an exponential potential.  Since near $\rho =0$ we have
\begin{equation}
K_{i \mu} (z) = \frac{1}{2} \left[ \Gamma(i \mu) \left( \frac{z}{2} \right)^{-i \mu} +
\Gamma(-i \mu) \left( \frac{z}{2} \right)^{i \mu} \right],
\end{equation}
we conclude that
\begin{equation}
\label{L2result} \int_{\rho > 0} \frac{d \rho}{\rho} \  K^*_{i
\mu_1}(m \rho) K_{i \mu_2}(m \rho) = \frac{\pi}{2} |\Gamma(i
\mu_1)|^2 \delta( \mu_1 - \mu_2).
\end{equation}

To study the Klein-Gordon factor on ${\cal H}$, we choose coordinates on ${\cal H}$ so that the line element takes the form
\begin{equation}
ds^2_{\cal H}  = - d\tau^2 + \cosh^2 (\tau) d \Omega^2_{d-2},
\end{equation}
where $d \Omega^2_{d-2}$ is the unit round metric on $S^{d-2}$.  We
take each harmonic to be of the form $Y_{q, \vec j} =  \exp(T_{q,
\vec j}(\tau)) Y_{\vec j}$, where $Y_{\vec j}$ are the standard
orthonormal harmonics on $S^{d-2}$.  Since $Y_{q, \vec j}$ satisfies
the massive wave equation on ${\cal H}$ we find for large $\tau$ (where the $|\vec j|^2$ term vanishes)
that
\begin{equation}
\left( \frac{d^2 T}{d \tau^2 }  + \frac{dT}{d\tau}\right)^2 + (d-2)
\frac{dT}{d\tau} + q^2 =0.
\end{equation}
This equation is easily solved to yield two solutions
\begin{equation}
\frac{dT}{d\tau} = - \frac{d-2}{2} \pm \sqrt{ \left( \frac{d-2}{2}
\right)^2 - q^2},
\end{equation}
where by convention we choose the $\pm$ sign to match the sign of
$-q$. The condition (\ref{Ynorm}) fixes $e^{2T(\tau =0)} =
\frac{2^{d-2}}{2\pi} q \left(q^2 - \left( \frac{d-2}{2} \right)^2
\right)^{-1/2} $. Taking $C$ in (\ref{KGHdef}) to lie in the distant future, it is
now straightforward to compute

\begin{eqnarray}
 \langle  Y^*_{-q, \vec j_1} ,  Y_{q, \vec j_2} \rangle_{KG, {\cal H}}
&=& 0, \cr
 \langle  Y^*_{q, \vec j_1} ,  Y_{q, \vec j_2} \rangle_{KG,
{\cal H}} &=&  \frac{q}{\pi}   \delta_{\vec j_1, \vec j_2}.
\end{eqnarray}
Here we have set $|q_1| = |q_2|$ as enforced by (\ref{L2result}).

The desired result is therefore
\begin{eqnarray}
\label{mnorms}
 \langle \tilde \phi^*_{q, \vec j},\phi_{q', \vec j'} \rangle_{KG}
 &:=& i \int_\Sigma \sqrt{g_\Sigma} \ \tilde \phi^*_{q, \vec j} n^a \overleftrightarrow \partial_a \phi_{q', \vec j'}
 = {\rm sign} (q) \frac{\mu }{2} |\Gamma(i\mu)|^2 \delta(q - q') \delta_{\vec j, \vec j'}.
\end{eqnarray}

\end{document}